\documentclass{elsart}

 \usepackage{graphics}

\usepackage{graphicx}
 \usepackage{epsfig}

\newcommand{\rr}{{\bf r}}
\newcommand{\vv}{{\bf v}}

\newcommand{\pp}{{\bf p}}
\newcommand{\uu}{{\bf u}}

\newcommand{\kk}{{\bf k}}
\newcommand{\jj}{{\bf j}}

\begin{document}

\begin{frontmatter}

\title{Irreversibility in the short memory approximation}

\author[label1]{Iliya V.\ Karlin}
\thanks[label1]{Corresponding author. ETH-Zentrum, Department of Materials,
Institute of Polymers, Sonneggstr. 3, ML J19, CH-8092 Z{\"u}rich, Switzerland Fax: +41 1 632 10 76 }
\ead{ikarlin@mat.ethz.ch}
\address{Department of Materials, Institute of Polymers\\
ETH Z{\"u}rich, CH-8092 Z{\"u}rich, Switzerland}

\author{Larisa L.\ Tatarinova}
\address{Department of Materials, Institute of Polymers\\
ETH Z{\"u}rich, CH-8092 Z{\"u}rich, Switzerland} \ead{larisa@iph.krasn.ru}
\address{Kirensky Institute of Physics SB RAS, Krasnoyarsk 660036, Russia}

\author{Alexander N.\ Gorban}
\ead{gorban@icm.krasn.ru}
\address{Institute of Computational Modeling SB RAS, Krasnoyarsk 660036,
Russia}

\author{Hans Christian {\"O}ttinger}
\ead{hco@mat.ethz.ch}
\address{Department of Materials, Institute of Polymers\\
ETH Z{\"u}rich, CH-8092 Z{\"u}rich, Switzerland}



\begin{abstract}
A recently introduced systematic approach to derivations of the macroscopic dynamics from the
underlying microscopic equations of motions in the short-memory approximation [Gorban et al, {\it
Phys. Rev. E}  {\bf 63}, 066124 (2001)] is presented in detail. The essence of this method is a
consistent implementation of Ehrenfest's idea of coarse-graining, realized via a matched expansion of
both the microscopic and the macroscopic motions. Applications of this method to a derivation of the
nonlinear Vlasov-Fokker-Planck equation, diffusion equation and hydrodynamic equations of the fluid
with a long-range mean field interaction are presented in full detail. The advantage of the method is
illustrated by the computation of the post-Navier-Stokes approximation of the hydrodynamics which is
shown to be stable unlike the Burnett hydrodynamics.

\end{abstract}


\end{frontmatter}

\section{Introduction}

The question of how irreversibility can be derived from reversible dynamics is one of the classical
problems in physics. The first solution has been suggested by Boltzmann \cite{Bol}, and it provoked
much discussion at that time. An alternative approach has been given by Ehrenfest \cite{Ehr} who
coined the notion of coarse-graining.

The impact of Ehrenfest's ideas on the long-standing discussions of the foundations of the
nonequilibrium thermodynamics is enormous (see, e.\ g.\  \cite{See,Rio}). In a recent  paper
\cite{GKOT} we have given a novel formalization of Ehrenfest's approach. The main focus of Ref.\
\cite{GKOT} was  the mathematical consistency of the formalization, whereas applications were only
briefly indicated. The goal of the present paper is to give a detailed description of the method,
focusing on how to apply it to various typical examples.

The starting point of our construction are microscopic equations of motion. A traditional example  of
the microscopic description is the  Liouville equation for classical particles. However, we need to
stress that the distinction between ``micro'' and ``macro'' is always context dependent. For example,
Vlasov's equation describes the dynamics of the one-particle distribution function. In one statement
of the problem, this is a microscopic dynamics in comparison to the evolution of hydrodynamic moments
of the distribution function. In a different setting, this equation itself  is a result of reducing
the description from the microscopic Liouville equation.

The problem of reducing the description includes a definition of the microscopic dynamics, and of the
macroscopic variables of interest, for which equations of the reduced description must be found. The
next step is the construction of the initial approximation. This is the well known quasi-equilibrium
approximation, which is the solution to the variational problem, $S\to {\rm max}$, where $S$ in the
entropy, under given constraints. This solution assumes that the microscopic distribution functions
depend on time only through their dependence on the  macroscopic variables. Direct substitution of
the quasi-equilibrium distribution function  into the microscopic equation of motion gives the
initial  approximation to  the macroscopic dynamics. All further corrections can be obtained from a
more precise approximation of the microscopic as well as of the macroscopic trajectories within a
given time interval $\tau$ which is the parameter of our  method.

The method described here has several clear advantages:

(i)  It allows to  derive complicated macroscopic equations, instead of writing them {\it ad hoc}.
This fact is especially significant for the description of complex fluids. The method gives explicit
expressions for relevant variables with one unknown parameter ($\tau$). This parameter can be
obtained from the experimental data.

(ii)  Another advantage of the method is its simplicity. For example, in the case where the
microscopic  dynamics is given by the Boltzmann equation, the approach avoids evaluation of Boltzmann
collision integral.

(iii) The most significant advantage of this formalization is that it is applicable to nonlinear
systems. Usually, in the classical approaches to reduced description, the  microscopic equation of
motion is linear. In that case, one can formally write the evolution operator in the exponential
form. Obviously, this does not work for nonlinear systems, such as, for example, systems with mean
field interactions. The method which we are presenting here is based on mapping the expanded
microscopic trajectory into the consistently expanded macroscopic trajectory. This does not require
linearity. Moreover, the order-by-order recurrent construction can be, in principle, enhanced by
restoring to other types of approximations, like Pad\'e approximation, for example, but we do not
consider these options here.

In the present paper we discuss in detail applications of the method \cite{GKOT} to  derivations of
macroscopic equations in various cases, with and without mean field interaction potentials, for
various choices of macroscopic variables, and demonstrate how computations are performed in the
higher orders of the expansion. The structure of the paper is as follows: In section 2,  for the sake
of completeness, we describe briefly the formalization of Ehrenfest's approach \cite{GKOT}. We stress
the r\^{o}le of the quasi-equilibrium approximation  as the starting point for the constructions to
follow. We derive explicit expressions for the correction to the quasi-equilibrium dynamics, and
conclude this section with the entropy production formula and its discussion. In section 3, we begin
the discussion  of applications. The first example is the derivation of the Fokker-Planck equation
from the Liouville equation. In section 4 we use the present formalism in order to derive
hydrodynamic equations. Zeroth approximation of the scheme is the  Euler equations of the
compressible nonviscous fluid. The first approximation leads to the system of Navier-Stokes
equations. Moreover, the approach allows to obtain the next correction, so-called post-Navier-Stokes
equations. The latter example is of particular interest. Indeed, it is well known that
post-Navier-Stokes equations as derived from the Boltzmann kinetic equation by the Chapman-Enskog
method (Burnett and super-Burnett hydrodynamics) suffer from unphysical instability already in the
linear approximation \cite{Boby}. We demonstrate it by the explicit computation that the linearized
higher-order hydrodynamic equations derived within our method are free from this drawback. In section
5, we derive macroscopic equations  in the  case of the nonlinear microscopic dynamics (hydrodynamic
equations from the Vlasov kinetic equation) . Significance of this example is that methods based on
the projection operator approach
 \cite{Zub,Rob,Grab},  are inapplicable to nonlinear systems.
We show in detail  how this problem is  solved on the basis of our method.

\section{General construction}

Let us consider a microscopic dynamics given by an equation,
\begin{equation}
\label{DYN} \dot{f}=J(f),
\end{equation}
where $f(x,t)$ is a distribution function over the phase space $x$ at  time $t$, and where operator
$J(f)$ may be linear or nonlinear. We consider linear macroscopic variables $M_k=\mu_k(f)$, where
operator $\mu_k$ maps $f$ into $M_k$. The problem is to obtain closed macroscopic equations of
motion, $\dot {M_k}=\phi_k(M)$. This is achieved in two steps: First, we construct an initial
approximation to the macroscopic dynamics and, second, this approximation is further corrected on the
basis of the coarse-gaining.

The initial  approximation is the quasi-equilibrium approximation, and it is based on the entropy
maximum principle under fixed constraints \cite{Jan,Gor}:
\begin{equation}
\label{SMAX} S(f)\to {\rm max},\ \mu(f)=M,
\end{equation}
where $S$ is the entropy functional, which is assumed to be strictly concave, and $M$ is the set of
the macroscopic variables $\{M\},$ and $\mu$ is the set of the corresponding operators. If the
solution to the problem (\ref{SMAX}) exists, it is unique thanks to the concavity of the entropy
functionals. Solution to equation (\ref{SMAX}) is called the quasi-equilibrium state, and it will be
denoted as $f^*(M)$. The classical example is the local equilibrium of the ideal gas: $f$ is the
one-body distribution function, $S$ is the Boltzmann entropy, $\mu$ are five linear operators,
$\mu(f)=\int\{1,{\vv},v^2\}fd{\vv}$, with ${\vv}$ the particle's velocity; the corresponding $f^*(M)$
is called the local Maxwell distribution function.

If the microscopic dynamics is given by equation (\ref{DYN}), then the quasi-equilibrium dynamics of
the variables $M$ reads:
\begin{equation}
\label{QEdyn} \dot{M}_k=  \mu_k(J(f^*(M))=\phi^*_k.
\end{equation}

The quasi-equilibrium approximation has  important property, it conserves the type of the dynamics:
If the entropy monotonically increases (or not decreases) due to equation (\ref{DYN}), then the same
is true for the quasi-equilibrium entropy, $S^*(M)=S(f^*(M))$, due to the quasi-equilibrium dynamics
(\ref{QEdyn}). That is, if
\[\dot{S}={\partial S(f)\over \partial f}\dot{f}={\partial S(f)\over
\partial f}J(f)\ge0,\]
then
\begin{equation}
\label{proof} \dot{S}^*=\sum_k\frac{\partial S^*}{\partial M_k}\dot{M}_k =\sum_k\frac{\partial
S^*}{\partial M_k} \mu_k(J(f^*(M)))\ge0.
\end{equation}
Summation in $k$ always implies summation or integration over the set of labels of the macroscopic
variables.

Conservation of the type of dynamics by the quasi-equilibrium approximation is a simple yet a general
and useful fact. If the entropy $S$ is an integral of motion of  equation (\ref{DYN}) then $S^*(M)$
is the integral of motion for the quasi-equilibrium equation (\ref{QEdyn}). Consequently, if we start
with a system which conserves the entropy (for example, with the Liouville equation) then we end up
with the quasi-equilibrium  system which conserves the quasi-equilibrium entropy. For instance, if
$M$ is the one-body distribution function, and (\ref{DYN}) is the (reversible)
 Liouville equation, then (\ref{QEdyn}) is the Vlasov
equation which  is reversible, too. On the other hand, if the entropy was monotonically increasing on
solutions to equation (\ref{DYN}), then the quasi-equilibrium entropy also increases monotonically on
solutions to the quasi-equilibrium dynamic equations (\ref{QEdyn}). For instance, if equation
(\ref{DYN}) is the Boltzmann equation for the one-body distribution function, and $M$ is a finite set
of moments (chosen in such a way that the solution to the problem (\ref{SMAX}) exists), then
(\ref{QEdyn}) are closed moment equations for $M$ which increase the quasi-equilibrium entropy (this
is the essence of a well known generalization of Grad's moment method).

\subsection{Enhancement of quasi-equilibrium approximations for
entropy-conserving dynamics}

The goal of the present  section is to describe the simplest analytic implementation,  the
microscopic motion with periodic coarse-graining. The notion of coarse-graining was introduced by P.\
and T.\ Ehrenfest's in their seminal work \cite{Ehr}: The phase space is partitioned into cells, the
coarse-grained variables are the amounts of the phase density inside the cells. Dynamics is described
by the two processes, by the Liouville equation for $f$, and by periodic coarse-graining, replacement
of $f(x)$ in each cell by its average value in this cell. The coarse-graining operation means
forgetting the microscopic details, or of the history.

\begin{figure}
\includegraphics{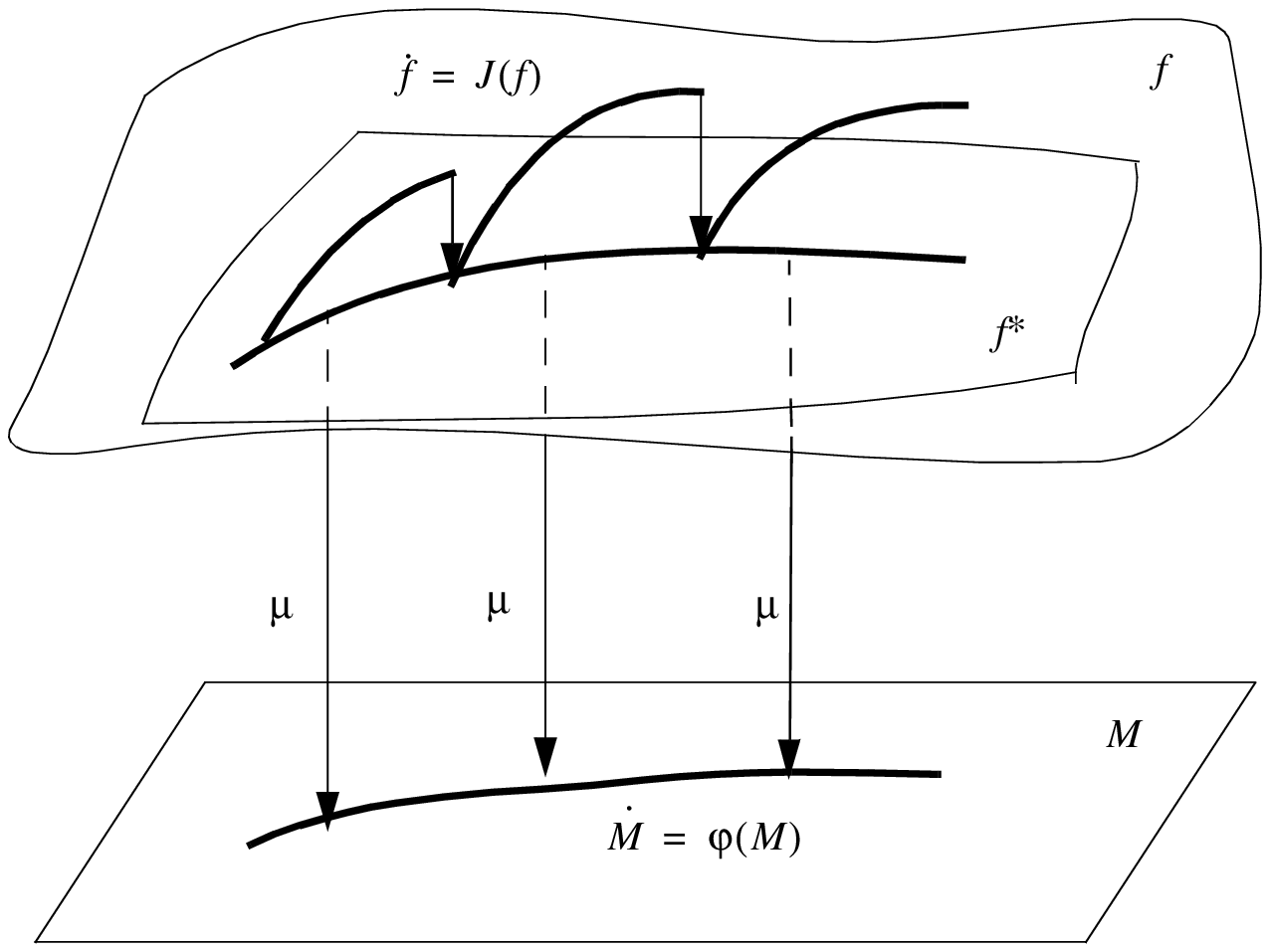}

\caption{Coarse-graining scheme. $f$ is the space of microscopic variables, $M$ is the space of the
macroscopic variables, $f^*$ is the quasi-equilibrium manifold, $\mu$ is the mapping from the
microscopic to the macroscopic space.} \label{Fig1}
\end{figure}

From the perspective of general quasi-equilibrium approximations, periodic coarse-graining amounts to
the return of the true microscopic trajectory on the quasi-equilibrium manifold with the preservation
of the macroscopic variables. The motion starts at the quasi-equilibrium state $f^*_i$. Then the true
solution $f_i(t)$ of the microscopic equation (\ref{DYN})
 with the initial condition $f_i(0)=f^*_i$
is coarse-grained  at a fixed time $t=\tau$, solution  $f_i(\tau)$ is replaced by the
quasi-equilibrium function $f^*_{i+1}=f^*(\mu(f_i(\tau)))$. This process is sketched in Fig.\
(\ref{Fig1}).

From the features of the quasi-equilibrium approximation it follows that for the motion with periodic
coarse-graining, the inequality is valid,
\begin{equation}
\label{shaking} S(f_i^*)\le S(f^*_{i+1}),
\end{equation}
the equality occurs if and only if the quasi-equilibrium is the invariant manifold of the dynamic
system (\ref{DYN}). Whenever the quasi-equilibrium is {\it not} the solution to equation (\ref{DYN}),
the strict inequality in (\ref{shaking}) demonstrates the entropy increase.

In other words, let us assume that the trajectory begins at the quasi-equilibrium manifold, then it
takes off from  this manifold according to the  microscopic evolution equations. Then, after some
time $\tau$, the trajectory is coarse-grained, that is the, state is brought back on the
quasi-equilibrium manifold keeping the values of the macroscopic variables. The irreversibility is
born in the latter process, and this construction clearly rules out quasi-equilibrium manifolds which
are invariant with respect to the microscopic dynamics, as candidates for a coarse-graining. The
coarse-graining indicates the way to derive equations for macroscopic variables from the condition
that the macroscopic trajectory, $M(t)$,  which governs the motion of the quasi-equilibrium states,
$f^*(M(t))$, should match precisely the same points on the quasi-equilibrium manifold,
$f^*(M(t+\tau))$, and this matching should be independent of both the initial time, $t$, and the
initial condition $M(t)$. The problem is then how to derive the continuous time macroscopic dynamics
which would be consistent with this picture. The simplest realization suggested in the Ref.\
\cite{GKOT} is based on using  an expansion of both the microscopic and the macroscopic trajectories.
Here we present this construction to the third order accuracy, in a general form, whereas only the
second-order accurate construction has been discussed in \cite{GKOT}.

Let us write down the solution to the microscopic equation (\ref{DYN}), and approximate this solution
by the polynomial of third oder in $\tau$. Introducing notation, $J^*=J(f^*(M(t)))$, we write,
\begin{equation}
f(t+\tau)=f^*+\tau J^*+{\tau^2\over 2}{\partial J^*\over \partial f} J^*+{\tau^3\over
3!}\left({\partial J^*\over \partial f}{\partial J^*\over \partial f} J^*+{\partial^2 J^*\over
\partial f^2}J^* J^*\right)+ o(\tau^3).\label{mic}
\end{equation}

Evaluation of the macroscopic variables on the function (\ref{mic}) gives
\begin{eqnarray}
M_k(t+\tau )&=& M_k+\tau \phi_k^*+{\tau^2\over 2}\mu_k\left({\partial J^*\over
\partial f} J^*\right)\nonumber\\* &+& {\tau^3\over 3!}\left\{\mu_k\left({\partial
J^*\over \partial f} {\partial J^*\over \partial f} J^*\right)+ \mu_k\left({\partial^2 J^*\over
\partial f^2} J^* J^*\right)\right\}+ o(\tau^3),\label{mic1}
\end{eqnarray}
where $\phi^*_k=\mu_k(J^*)$ is the quasi-equilibrium macroscopic vector field (the right hand side of
equation (\ref{QEdyn})), and all the functions and derivatives are taken in the quasi-equilibrium
state at time $t$.

We shall now establish the macroscopic dynamic by matching the macroscopic and the microscopic
dynamics. Specifically, the macroscopic dynamic equations (\ref{QEdyn}) with the right-hand side not
yet defined, give the following third-order result:
\begin{eqnarray}
M_k(t+\tau)&=&M_k+\tau\phi_k+{\tau^2\over 2}\sum_j {\partial \phi_k\over
\partial M_j}\phi_j \nonumber\\*&+&{\tau^3\over 3!} \sum_{ij}\left(
{\partial^2\phi_k\over
\partial M_iM_j}\phi_i\phi_j +{\partial \phi_k\over \partial
M_i}{\partial \phi_i\over \partial M_j}\phi_j \right)+o(\tau^3).\label{mak}
\end{eqnarray}

Expanding functions $\phi_k$ into the series $\phi_k=R_k^{(0)}+\tau R_k^{(1)}+\tau^2R_k^{(2)}+...$,
($R_k^{(0)}=\phi^*$),  and requiring that the microscopic and the macroscopic dynamics coincide to
the order of $\tau^3$, we obtain the sequence of corrections for the right-hand side of the equation
for the macroscopic variables. Zeroth order is the quasi-equilibrium approximation to the macroscopic
dynamics. The first-order correction gives:
\begin{eqnarray}
R_k^{(1)}={1\over 2}\left\{\mu_k\left({\partial J^*\over \partial f} J^*\right)-\sum_j {\partial
\phi_k^*\over
\partial M_j}\phi_j^*\right\}\label{Rk1}.
\end{eqnarray}
The next, second-order correction has the following explicit form:
\begin{eqnarray}
R_k^{(2)}&=&{1\over 3!} \left\{\mu_k \left({\partial J^*\over
\partial f}{\partial J^*\over \partial f} J^*\right) + \mu_k\left({\partial^2
J^*\over
\partial f^2}J^* J^*\right)\right\} - {1\over 3!} \sum_{ij} \left( {\partial
\phi_k^*\over \partial M_i} {\partial \phi_i^*\over \partial M_j}\phi_j^*\right)\nonumber\\*
&-&{1\over 3!}\sum_{ij}\left({\partial^2 \phi_k^*\over
\partial M_i\partial M_j} \phi_i^*\phi_j^* \right)-{1 \over 2 }
\sum_j \left( {\partial \phi_k^*\over \partial M_j} R_j^{(1)}+ {\partial R_j^{(1)}\over \partial M_j}
\phi_j^*\right),\label{26}
\end{eqnarray}
Further corrections are found by the same token. Equations (\ref{Rk1})--(\ref{26})  give explicit
closed expressions for corrections to the quasi-equilibrium dynamics to the order of accuracy
specified above. They are used below in various specific examples.

\subsection{Entropy production}

The most important consequence of the above construction is that the resulting continuous time
macroscopic equations retain the dissipation property of the discrete time coarse-graining
(\ref{shaking}) on each order of approximation $n\ge 1$. Let us first consider the entropy production
formula for the first-order approximation. In order to shorten notations, it is convenient to
introduce the quasi-equilibrium projection operator,

\begin{equation}
P^*g= \sum_k{\partial f^*\over
\partial M_k}\mu_k(g).\end{equation}
It has been demonstrated in \cite{GKOT} that the entropy production,

\[\dot{S}^*_{(1)}=\sum_k\frac{\partial S^*}{\partial M_k}(R_k^{(0)}+\tau
R_k^{(1)}),\] equals
\begin{eqnarray}
\dot{S}^*_{(1)} =-\frac{\tau}{2}(1-P^*)J^*\left.{\partial ^2 S^*\over
\partial f
\partial f
}\right|_{f^*}(1-P^*)J^*.\label{RESULT2}
\end{eqnarray}
Equation (\ref{RESULT2}) is nonnegative definite due to concavity of the entropy. Entropy production
(\ref{RESULT2}) is equal to zero only if the quasi-equilibrium approximation is the true solution to
the microscopic dynamics, that is, if $(1-P^*)J^*\equiv0$. While quasi-equilibrium approximations
which solve the Liouville equation are uninteresting objects (except, of course, for the equilibrium
itself), vanishing of the entropy production in this case is a simple test of consistency of the
theory. Note that the entropy production (\ref{RESULT2}) is proportional to $\tau$. Note also that
projection operator does not appear in our consideration a priory, rather, it is the result of
exploring the coarse-graining condition in the previous section.

Though equation (\ref{RESULT2}) looks very natural, its existence is rather subtle. Indeed, equation
(\ref{RESULT2}) is a difference of the two terms, $\sum_k\mu_k( J^*\partial J^*/\partial f)$
(contribution of the second-order approximation to the microscopic trajectory), and
$\sum_{ik}R_i^{(0)}\partial R_k^{(0)}/\partial M_i$ (contribution of the derivative of the
quasi-equilibrium vector field). Each of these expressions separately gives a positive contribution
to the entropy production, and equation (\ref{RESULT2}) is the difference of the two positive
definite expressions. In the higher order approximations, these subtractions are more involved, and
explicit demonstration of the entropy production formulae becomes a formidable task. Yet, it is
possible to demonstrate the increase-in-entropy without explicit computation, though at a price of
smallness of $\tau$. Indeed, let us denote $\dot{S}^*_{(n)}$ the time derivative of the entropy on
the $n$th order approximation. Then

\begin{eqnarray}
\int_t^{t+\tau}\dot{S}^*_{(n)}(s)ds=S^*(t+\tau)-S^*(t)+O(\tau^{n+1}),\nonumber
\end{eqnarray}

where $S^*(t+\tau)$ and $S^*(t)$ are true values of the entropy at the adjacent states of the
$H$-curve. The difference $\delta S=S^*(t+\tau)-S^*(t)$ is strictly positive for any fixed $\tau$,
and, by equation (\ref{RESULT2}), $\delta S\sim \tau^2$ for small $\tau$. Therefore, if $\tau$ is
small enough, the right hand side in the above expression is positive, and $$\tau
\dot{S}_{(n)}^*(\theta_{(n)})>0,$$ where $t\le\theta_{(n)}\le t+\tau$. Finally, since
$\dot{S}_{(n)}^*(t)=\dot{S}_{(n)}^*(s)+O(\tau^n)$ for any $s$ on the segment $[t,t+\tau]$, we can
replace $\dot{S}^*_{(n)}(\theta_{(n)})$ in the latter inequality by $\dot{S}^*_{(n)}(t)$. The sense
of this consideration is as follows: Since the entropy production formula (\ref{RESULT2}) is valid in
the leading order of the construction, the entropy production will not collapse in the higher orders
at least if the coarse-graining time is small enough. More refined estimations can be obtained only
from the explicit analysis of the higher-order corrections.

\subsection{Relation to the work of Lewis}

Among various realizations of the coarse-graining procedures, the work of Lewis \cite{Lewis} appears
to be most close to our approach. It is therefore pertinent to discuss the differences. Both methods
are based on the coarse-graining condition,
\begin{equation}\label{CG}
  M_k(t+\tau)=\mu_k\left(T_{\tau}f^*(M(t))\right),
\end{equation}
where $T_{\tau}$ is the formal solution operator of the microscopic dynamics. Above, we applied a
consistent expansion of both, the left hand side and the right hand side of the coarse-graining
condition (\ref{CG}), in terms of the coarse-graining time $\tau$. In the work of Lewis \cite{Lewis},
it was suggested, as a general way to exploring the condition (\ref{CG}), to write the first-order
equation for $M$ in the form of the differential pursuit,
\begin{equation}\label{Lewis}
  M_k(t)+\tau\frac{dM_k(t)}{dt}\approx\mu_k\left(T_{\tau}f^*(M(t))\right).
\end{equation}
In other words, in the work of Lewis \cite{Lewis}, the expansion to the first order was considered on
the left (macroscopic) side of equation (\ref{CG}), whereas the right hand side containing the
microscopic trajectory $T_{\tau}f^*(M(t))$ was not treated on the same footing. Clearly, expansion of
the right hand side to first order in $\tau$ is the only equation which is common in both approaches,
and this is the quasi-equilibrium dynamics. However, the difference occurs already in the next,
second-order term (see Ref.\ \cite{GKOT} for details). Namely, the expansion to the second order of
the right hand side of Lewis' equation (\ref{Lewis}) results in a dissipative equation (in the case
of the Liouville equation, for example) which remains dissipative even if the quasi-equilibrium
approximation is the exact solution to the microscopic dynamics, that is, when microscopic
trajectories once started on the quasi-equilibrium manifold belong to it in all the later times, and
thus no dissipation can be born by any coarse-graining.

On the other hand, our approach assumes a certain smoothness of trajectories so that application of
the low-order expansion bears physical significance. For example, while using lower-order truncations
it is not possible to derive the Boltzmann equation because in that case the relevant
quasi-equilibrium manifold ($N$-body distribution function is proportional to the product of one-body
distributions, or uncorrelated states, see next section) is almost invariant during the long time (of
the order of the mean free flight of particles), while the trajectory steeply leaves this manifold
during the short-time pair collision. It is clear that in such a case lower-order expansions of the
microscopic trajectory do not lead to useful results. It has been clearly stated by Lewis
\cite{Lewis}, that the exploration of the condition (\ref{CG}) depends on the physical situation, and
how one makes approximations. In fact, derivation of the Boltzmann equation given by Lewis on the
basis of the condition (\ref{CG}) does not follow the differential pursuit approximation: As is well
known, the expansion in terms of particle's density of the solution to the BBGKY hierarchy is
singular, and begins with the \textit{linear} in time term. Assuming the quasi-equilibrium
approximation for the $N$-body distribution function under fixed one-body distribution function, and
that collisions are well localized in space and time, one gets on the right hand side of equation
(\ref{CG}),
\[f(t+\tau)=f(t)+n\tau J_B(f(t))+o(n),\]
where $n$ is particle's density, $f$ is the one-particle distribution function, and $J_B$ is the
Boltzmann's collision integral. Next, using the mean-value theorem on the left hand side of the
equation (\ref{CG}), the Boltzmann equation is derived (see also a recent elegant
renormalization-group argument for this derivation \cite{RG}).

We stress that our approach of matched expansion for exploring the coarse-graining condition
(\ref{CG}) is, in fact, the exact (formal) statement that the unknown macroscopic dynamics which
causes the shift of $M_k$ on the left hand side of equation (\ref{CG}) can be reconstructed
order-by-order to any degree of accuracy, whereas the low-order truncations may be useful for certain
physical situations. A thorough study of the cases beyond the lower-order truncations is of great
importance which is left for future work.

\section{Vlasov-Fokker-Planck kinetic equation}\label{VFPS}

In this section we derive kinetic equations based on the approach formulated above. Here microscopic
dynamics is given by the $N-$particle Liouville equation. Macroscopic variable is the one-particle
distribution function. The solution to the variational problem (\ref{SMAX}) is the approximation of
the absence of correlations. In this case the quasi-equilibrium $N$-particle distribution function is
proportional to the product of the one-particle distribution functions. On the basis of this
quasi-equilibrium we obtain the Vlasov equation, as the zeroth approximation, and the Fokker-Planck
equation, as the next correction.

The dynamics of the ensemble of $N$ classical point particles interacting by a pair potential is
given by the Liouville equation,
\begin{eqnarray}{\partial {\it w}_N\over \partial t}&=&L{\it w}_N=\{H,{\it
w}_N\}\nonumber\\*&=&\sum_{ij}\left( {\partial \Phi(|\rr_i-\rr_j|)\over
\partial \rr_i}{\partial {\it w}_N\over
\partial \pp_i}+ {\partial U(\rr_i)\over
\partial \rr_i}{\partial {\it w}_N\over \partial \pp_i}-{\pp_i\over
m}{\partial {\it w}_N\over \partial \rr_i}\right),\label{le11}
\end{eqnarray}
where $w_N$ is the $N-$particle distribution function,
\begin{eqnarray}w_N=w_N(\pp_1,...\pp_N,\rr_1,...\rr_N,t).\nonumber
\end{eqnarray}
 It is normalized to one ($\int w_N d\pp_1...d\rr_N=1$).
$L$ is the Liouville operator (Poisson bracket), $H$ is classical Hamilton function,
$\Phi(|\rr_i-\rr_j|)$ describes  interaction between particles with indices $i$ and $j$, $U(\rr_i)$
is an external field. Here, and in every case later, summation is assumed over the total number of
particles. The indices $i$ and $j$ take values from $1$ to $N$ independently.

The macroscopic variable is the one-particle distribution function $f(\pp,\rr,t)$. The latter
satisfies the normalization condition $\int f(\pp,\rr,t)d\rr d\pp =N$.  The mapping of the
microscopic variables into the space of macroscopic variables is given by the following operator:
\begin{eqnarray}
\mu=\sum_i\int \delta(x-y_i)dy_1...dy_N,\label{map}
\end{eqnarray}
where  $x$ denotes the point $(\pp,\rr)$ in the phase space. The solution to the variational problem
(\ref{SMAX}) has the form:
\begin{eqnarray}{\it
w^*}_N(x_1,...,x_N,t)=N^{-N}\prod_if(x_i,t),\label{w1}
\end{eqnarray}

Before proceeding further, a remark on the choice of the one-particle distribution function as the
macroscopic variable is in order. The total energy of the system under consideration can be expressed
as a linear functional of the \textit{two}-particle distribution function rather than in terms of the
one-particle distribution. For that reason, when only the one-particle distribution is chosen for the
macroscopic variable, the coarse-graining procedure must be supplemented by a termostatting procedure
in order to keep the energy balance in the system intact. While it is possible to take care of the
energy conservation through introducing extra terms into the Liouville equation describing
interactions with the thermostat, we will implement thermostatting after the coarse-graining. Now we
proceed with executing the formalism developed in the previous sections.

The quasi-equilibrium (conservative) dynamics in the space of the macroscopic variables is given by
the action of the operator (\ref{map}) on the microscopic equation of motion (\ref{le11}).
\begin{eqnarray}
&{ }&{\partial f(x,t)\over \partial t}= \sum_i\int \delta (x-y_i)dy_1...dy_N\nonumber\\*&\times &
\sum _k\left( {\partial U(\eta_k)\over
\partial \eta_k}{\partial {\it w}_N^*\over \partial \xi_k}-{\pp_k\over
m}{\partial {\it w}_N^* \over \partial \eta_k }+ \sum _{l\neq k}{
\partial \Phi (|\eta_k-\eta_l|)\over \partial \eta_k }{\partial
{\it w}_N^*\over
\partial \xi_k }\right).\nonumber
\end{eqnarray}
In order to avoid a confusion, we used variables $\eta,$ $\xi,$ and $y$ under integrals instead of
$\rr,$ $\pp,$ and $x$, respectively. Thus, we obtain:
\begin{eqnarray}
{\partial f(x,t) \over \partial t}+{{\pp }\over m}{\partial f(x,t)\over
\partial \rr}-{\partial U(\rr)\over \partial \rr}{\partial
f(x,t) \over
\partial \pp} +F(\rr,t){\partial
f(x,t) \over
\partial \pp}=0, \label{V1}
\end{eqnarray}
where
\begin{eqnarray}
F(\rr,t)&=&\int {\partial \Phi(|\rr-\eta|)\over
\partial \rr}f(y,t)dy\label{FF}
\end{eqnarray}
is the mean field interaction force. Equation (\ref{V1}) has been first derived by Vlasov, and is
usually applied to description of plasma without collisions \cite{Lan,Res}.

Now we are going to obtain dissipative correction to the Vlasov equation (\ref{V1}) based on the
approach developed above. For the sake of simplicity we omit external field term. Let us begin with
the second term in the expression (\ref{Rk1}).
\begin{eqnarray}
{\partial \phi^*\over \partial M}\phi^*&=&\left( {\pp\over m}\right)^2 {\partial \over
\partial \rr}\left({\partial f\over \partial
\rr}\right)\label{rs}\\*&+&{\pp\over m}{\partial \over
\partial \rr}\left({\partial f\over \partial
\pp}F(\rr)\right)+{\partial f\over
\partial
\pp}\Psi(\rr,t)\nonumber\\*&+&F(\rr,t){\partial \over
\partial \pp}\left({\pp\over m}{\partial f\over \partial
\rr}\right)+F(\rr,t){\partial \over \partial \pp}\left({\partial f\over \partial
\pp}F(\rr,t)\right),\nonumber
\end{eqnarray}
where
\begin{eqnarray}
\Psi(\rr,t)&=&{1\over m}\int {\partial \Phi(|\rr-\rr'|)\over
\partial \rr}\left(\vv'{\partial f(x',t)\over
\partial \rr'}\right) dx'.\label{Pmf}
\end{eqnarray}

First term in equation  (\ref{Rk1}) is proportional to:
\begin{eqnarray}
{\partial J^*\over \partial f}J^*&=&L^2w^*_N\nonumber\\* &=&\left[\sum_i-{\pp_i\over m}{\partial
\over
\partial \rr_i }+\sum_{i,j\neq i}{\partial \Phi(|\rr_i-\rr_j|)\over
\partial \rr_i}
{\partial \over \partial \pp_i}\right]\nonumber\\* &\times & \left[\sum_k-{\pp_k\over m}{\partial
\over
\partial \rr_k }+\sum_{k,l\neq i}{\partial \Phi(|\rr_k-\rr_l|)\over
\partial \rr_k}
{\partial \over \partial \pp_k} \right]\nonumber\\*&\times&{\prod_sf(x_s,t)} N^{-N},\label{fL}
\end{eqnarray}
where indices  $i,$ $j,$ $k,$ and $l$ enumerate particles in the system.

Removing the brackets in (\ref{fL}), we obtain,
\begin{eqnarray}
&&\left[\sum_{ik}{\pp_i\pp_k\over m^2}{\partial^2\over
\partial \rr_i
\partial\rr_k}-2\sum_{i,l,k}{\pp_i\over m}{\partial
\Phi(|\rr_k-\rr_l|)\over
\partial\rr_k}{\partial^2\over \partial \rr_i\partial
\pp_k}-\sum_{ij}{1\over m}{\partial \Phi(|\rr_i-\rr_j|)\over \partial\rr_i}{\partial \over
\partial \pp_i}+\right.\nonumber\\*&&\left.\sum_{ijkl}{\partial
\Phi(|\rr_i-\rr_j|)\over \partial\rr_k} {\partial \Phi(|\rr_k-\rr_l|)\over
\partial\rr_k}{\partial^2\over
\partial \pp_i \partial
\pp_k}\right]{\prod_s f(x_s,t)}N^{-N},\label{long}
\end{eqnarray}
where $j\neq i$ and $k \neq l.$

Let us now consider the first term in (\ref{long}) under the action of the operator (\ref{map}). The
result is not equal to zero only for $i=k$. In this case we have:
\begin{eqnarray}
\left({\pp\over m}\right)^2 {\partial \over
\partial \rr}\left({\partial f\over \partial \rr}\right).\nonumber
\end{eqnarray}
Note that this term is cancelled by the same term in equation (\ref{rs}). The second and the third
term in  equation (\ref{long}) do not contribute to the final result by the same reason.

Let us consider the last term in (\ref{long}), whose contribution is nontrivial. It is:
\begin{eqnarray}
&&\sum_{p}\int \delta(x-y_p)\sum_{ijkl}{\partial \Phi(|\rr_i-\rr_j|)\over
\partial\rr_i}{\partial \Phi(|\rr_k-\rr_l|)\over
\partial\rr_k}\nonumber\\*
&\times& {\partial^2\over \partial \pp_i
\partial \pp_k}{\prod_{s}f(y_s,t)} N^{-N}dy_1..dy_N,\label{lt}
\end{eqnarray}
where $k\neq l$ and $i\neq j$.

i) For $j=l$, expression (\ref{lt}) gives:
\begin{eqnarray}
{\partial^2f\over \partial p^2}\int \left({\partial \Phi(|\rr-\rr'|)\over
\partial\rr}\right)^2f(x')dx'{N^{N-2}N(N-1)\over N^N}.\label{tl}
\end{eqnarray}
In the thermodynamic limit ($N \to \infty$) the expression (\ref{tl}) becomes equal to the following:
\begin{eqnarray}
{\partial^2f\over \partial p^2}\int \left({\partial \Phi(|\rr-\rr'|)\over
\partial\rr}\right)^2f(x')dx'.\label{tl1}
\end{eqnarray}

ii) For $j\neq l$, expression (\ref{lt}) gives:
\begin{eqnarray}
{\partial^2f\over \partial p^2} \left(\int{\partial \Phi(|\rr-\rr'|)\over
\partial\rr}f(x')dx'\right)^2{N^{N-3}(N-1)^2\over N^{N-1}}.\label{tl2}
\end{eqnarray}

Combining the expressions (\ref{tl1}) and (\ref{tl2}) with the last term in the equation (\ref{rs}),
we obtain:
\begin{eqnarray}
\left(-{1\over N}+{1\over N^2}\right){\partial^2f\over \partial p^2} \left(\int{\partial
\Phi(|\rr-\rr'|)\over
\partial\rr}f(x')dx'\right)^2.\nonumber
\end{eqnarray}
In the thermodynamic limit the term $1/N^2$ can  be neglected.

Thus, we obtain the following macroscopic equation of motion:
\begin{eqnarray}
&{ }&{\partial f(x,t)\over \partial t}+{\pp\over m}{\partial f(x,t)\over
\partial \rr}-{\partial f(x,t)\over \partial \pp}
\int {\partial \Phi(|\rr-\eta|)\over \partial \rr}f(y,t)dy=\nonumber\\&=&D{\partial^2 f(x,t)\over
\partial p^2},\label{Lan}
\end{eqnarray}
where  $D$ is the diffusion tensor,
\begin{eqnarray}
D={\tau\over 2}\left\{ \int \left({\partial \Phi(|\rr-\eta|)\over
\partial \rr}\right)^2f(y,t)dy-{1\over N}\left( \int{\partial
\Phi(|\rr-\eta|)\over
\partial \rr}f(y,t)dy\right)^2 \right\}.\label{D}
\end{eqnarray}
Note the clear Green-Kubo-type structure of the latter expression. The absence of the time
integration reveals the short-memory nature of the construction.

At this point of the derivation, one can notice that the obtained collision integral in equation
(\ref{Lan}) does not conserve the total energy. As we have argued above, this fact has the following
explanation: Since the interaction between particles is pair-wise, the total energy is a functional
of the two-particles distribution function. However, we have restricted ourselves to the equation for
the one-particle distribution function function. At the beginning of our procedure, the trajectories
belong to the quasi-equilibrium manifold, later the correlations start growing, and lead to the
decrease of the potential energy, and to the increase of the kinetic energy. At every instance of the
microscopic motion, the total energy is conserved. At some moment (defined by the step of
coarse-graining) the system is returned back onto the quasi-equilibrium manifold with the
conservation of value of all the variables which can be expressed as functionals of the one-particle
distribution function (including the kinetic energy). As a result, the total energy is not conserved.
There are two ways to solve this problem. The first is to choose the two-particle distribution
function as the  macroscopic variable. This route  is very complicated, and in order to circumvent it
and to stay on the level of the one-particle distribution function, we shall subtract the spurious
contribution by a regularization of the macroscopic vector field.

Specifically, let us write the dissipative contribution to equation (\ref{Lan}) in the gradient form:
\begin{eqnarray}
{\partial f\over \partial t}_{\textrm{diss}}=-{\partial \jj\over
\partial \pp},\label{gradi}
\end{eqnarray}
here \begin{eqnarray} \jj=-D{\partial f\over \partial \pp}.\nonumber \end{eqnarray} This form
automatically takes into account the conservation of particle's density. In order to satisfy the
conservation of the momentum and of the energy, we shall introduce a subspace $E$, and require that
$\jj$ belong to $E.$ Subspace $E$ is a set of functions $\{\varphi\}$ for which the momentum and the
energy are conserved. This means that the moments,
\begin{eqnarray}
M_0=\int\varphi d\pp,\hspace{1cm} M_{\alpha}=\int p_{\alpha} \varphi_{\alpha} d\pp,\nonumber
\end{eqnarray}
are equal to zero $(M_0=0,M_{\alpha}=0)$. Index $\alpha$ runs over the space coordinates. There is no
summation over $\alpha$.

Now we introduce an orthogonal complement to $E$. This is a set of functions $\phi_i$, ($i$ is $0,\
x,\ y,\ z$) which satisfy the conditions:
\begin{eqnarray}
(\varphi,\phi_i)&=&\int {1\over f}\varphi \phi_id\pp=0,\nonumber\\* (\phi_i,\phi_j)&=&\int {1\over
f}\phi_i\phi_jd\pp=\delta_{ij},\label{prod}
\end{eqnarray}
where we have introduced the scalar product $(\cdot ,\cdot)$ to be used below, and where
$\delta_{ij}$ is the  Kronecker delta, and $f$ is the distribution function. We define a projector
$\Pi$ which  maps $\jj$ into $\{E\}$. The projector is:
\begin{eqnarray}
\Pi(\jj)=\jj-\sum_i^4(\phi_i,\jj)\phi_i.\label{Pi}
\end{eqnarray}

The final form of the macroscopic equation depends on the choice of the functions $\phi_i$,
specifically, on the conditions of normalization and orthogonality. Let us demonstrate a few cases.

Let us first assume that the functions $\phi_i$  satisfy conditions, $(\phi_i,\phi_j)=\delta_{ij}$,
and that the average momentum of the fluid, $\int f \pp d\pp$, is  equal to zero. Then we can take
these functions in the form:
\begin{eqnarray}
\phi_0&=&C_0f,\nonumber\\ \phi_{\alpha}&=&C_{\alpha} p_{\alpha}f.\label{bas}
\end{eqnarray}
where the constants $C_i$ should be found from the normalization conditions
\begin{eqnarray}
C_0^2\int fd{\pp}&=&C_0^2n=1,\hspace{1.5cm} C_0=1/\sqrt{n},\nonumber\\* C_{\alpha}^2\int
p_{\alpha}^2fd\pp&=&C_{\alpha}^2mnk_BT=1,\hspace{0.7cm} C_{\alpha}={1\over \sqrt{mnk_BT}}.\nonumber
\end{eqnarray}

Now we calculate the convolutions $(\phi_i,\jj):$
\begin{eqnarray}
(\phi_0,\jj)&=&-D\int_{-\infty}^{+\infty}{\partial f(x,t)\over
\partial \pp}d\pp=0,\label{phi3,q}\\* (\phi_{\alpha},\jj)&=&-D{1\over
\sqrt{mnk_BT}} \int_{-\infty}^{+\infty} p_{\alpha}{\partial f(x,t)\over
\partial p_{\beta}}d\pp=\left\{ \begin{array}{cl} -Dn/\sqrt{mnk_BT}, & \mbox{if}\hspace{0.2cm}
 \alpha=\beta \\* 0, & \mbox{if}\hspace{0.2cm} \alpha \neq \beta.
                                 \end{array}                   \right.\nonumber
\end{eqnarray}

Finally, substituting  result (\ref{phi3,q}) into equation (\ref{Pi}) we obtain:
\begin{eqnarray}
\jj=-D\left({\partial f(x,t) \over \partial \pp}+{1\over mk_BT}{\pp} f(x,t)\right).\label{Pi'}
\end{eqnarray}
Thus, equation  (\ref{Lan}) is represented as follows:
\begin{eqnarray}
&{ }&{\partial f(x,t)\over \partial t}+{\pp\over m}{\partial f(x,t)\over
\partial \rr}-{\partial f(x,t)\over \partial \pp}
\int {\partial \Phi(|\rr-\eta|)\over \partial \rr}f(y,t)dy=\nonumber\\&=&D{\partial \over \partial
{\pp}}\left({\partial f(x,t) \over
\partial \pp}+{1\over mk_BT}{\pp} f(x,t)\right).\label{fin}
\end{eqnarray}
This  is the Fokker-Planck equation describing a diffusion in the momentum space. The short-memory
approximation behind the present approach is apparent in the present example.

The result (\ref{fin}) can be generalized to the case when the average momentum $\uu$ does not
vanish. In this case the basis functions (\ref{bas}) are chosen as follows:
\begin{eqnarray}
\phi_0&=&f/\sqrt{n},\nonumber\\ \phi_{\alpha}&=&{p_{\alpha}-mu_{\alpha}\over
\sqrt{nmk_BT}}f.\nonumber
\end{eqnarray}
It is easy to check that they satisfy the condition (\ref{prod}). Then equation (\ref{gradi}) reads,
\begin{eqnarray}
{\partial f(x,t)\over \partial t}_{\textrm{diss}}=D{\partial \over
\partial \pp}\left({\partial f(x,t) \over \partial \pp}+{{\pp} -
m{\uu }\over mk_BT}f(x,t)\right).\label{fin2}
\end{eqnarray}

Thus, in this section  we derived the diffusion equation from the $N$-particle Liouville equation. In
contrast to a commonly known Fokker-Planck equation, equation (\ref{fin}) is nonlinear because  the
diffusion coefficient depends on the distribution function. Near equilibrium, the diffusion
coefficient becomes constant, and the result is the usual Fokker-Planck equation (with a mean-field
extension due to Vlasov's term). Then the usual stochastic interpretation in terms of the
fluctuation-dissipation theorem applies. We remind the reader that the stochastic interpretation of
nonlinear equations is, in general, a difficult problem \cite{HCO}, and that the result obtained here
corresponds to the lower-order ($\tau^2$) of approximation of microscopic trajectories. Higher-order
corrections will cease to have a stochastic interpretation in terms of the usual
fluctuation-dissipation theorem.

\section{Equations of hydrodynamics for simple fluid}

The method discussed above enables one to establish in a simple way the form of equations of the
macroscopic dynamics to various degrees of approximation.  In this section, the microscopic dynamics
is given by the Liouville equation, similar to the previous case. However, we take another set of
macroscopic variables: density, average velocity, and average temperature of the fluid. Under this
condition the solution to the problem (\ref{SMAX}) is the local Maxwell distribution. For the
hydrodynamic equations, the zeroth (quasi-equilibrium) approximation is given by Euler's equations of
compressible nonviscous fluid. The next order approximation are the Navier-Stokes equations which
have dissipative terms.

Higher-order approximations to the hydrodynamic equations, when they are derived from the  Boltzmann
kinetic equation (so-called Burnett approximation), are subject to various difficulties, in
particular, they exhibit an  instability of sound waves at sufficiently short wave length (see, e.\
g. \cite{KG2002} for a recent review). Here we demonstrate how model hydrodynamic equations,
including post-Navier-Stokes approximations, can be derived on the basis of coarse-graining idea, and
investigate the linear  stability of the obtained equations. We will find that the resulting
equations are stable.

Two points need a clarification before we proceed further \cite{GKOT}. First, below we consider the
simplest Liouville equation for the one-particle distribution, describing a free moving particle
without interactions. The procedure of coarse-graining we use is an implementation of collisions
leading to dissipation. If we had used the full interacting $N$-particle Liouville equation, the
result would be different, in the first place, in the expression for the local equilibrium pressure.
Whereas in the present case we have the ideal gas pressure, in the $N$-particle case the non-ideal
gas pressure would arise.

Second, and more essential is that, to the order of the Navier-Stokes equations, the {\it result} of
our method is identical to the lowest-order Chapman-Enskog method as applied to the Boltzmann
equation with a single relaxation time model collision integral (the Bhatnagar-Gross-Krook model
\cite{BGK}). However, this happens only at this particular order of approximation, because already
the next, post-Navier-Stokes approximation,  is different from the Burnett hydrodynamics as derived
from the BGK model (the latter is linearly unstable).

\subsection{Derivation of the Navier-Stokes equations}\label{ChH}
Let us assume that reversible microscopic dynamics is given by the one-particle Liouville equation,
\begin{eqnarray}
{\partial f\over \partial t}=-v_i{\partial f\over \partial r_i},\label{li}
\end{eqnarray}
where $f=f(\rr,\vv,t)$ is the one-particle distribution function, and index $i$ runs over spatial
components $\{x,\ y,\ z \}$. Subject to appropriate boundary conditions which we assume, this
equation conserves the Boltzmann entropy $S=-k_{\rm B}\int f\ln f d\vv d\rr$.

We introduce the following hydrodynamic moments as the macroscopic variables: $M_0=\int
fd\vv,\,M_i=\int v_ifd\vv,\,M_4=\int v^2fd\vv $. These variables are related to the more conventional
density, average velocity and temperature, $n,\,\uu,\,T$ as follows:
\begin{eqnarray}
&&M_0=n,\hspace{0.5cm}M_i=nu_i,\hspace{0.7cm}M_4={3nk_BT\over m}+nu^2,\nonumber\\
&&n=M_0,\hspace{0.5cm}u_i=M_0^{-1}M_i,\hspace{0.5cm}T={m\over 3k_BM_0}(M_4-M_0^{-1}M_iM_i).\label{M}
\end{eqnarray}
The quasi-equilibrium distribution function  (local Maxwellian) reads:
\begin{eqnarray}
f_0=n\left({m\over 2\pi k_BT}\right)^{3/2}\exp\left({-m(v-u) ^2\over 2k_BT}\right).\label{QEM}
\end{eqnarray}
Here and below, $n,\ \uu,$ and $T$ depend on $\rr$ and $t.$

Based on the microscopic dynamics (\ref{li}), the set of macroscopic variables (\ref{M}), and the
quasi-equilibrium (\ref{QEM}), we can derive the  equations of the macroscopic motion.

A specific feature of the present example is that the quasi-equilibrium equation for the density (the
continuity equation),
\begin{eqnarray}
{\partial n\over
\partial t}&=&-{\partial nu_i\over \partial r_i},\label{n}
\end{eqnarray}
should be excluded out of the further corrections. This rule should be applied generally: If a part
of the chosen macroscopic variables (momentum flux $n\uu$ here) correspond to fluxes of other
macroscopic variables, then the quasi-equilibrium equation for the latter is already exact, and has
to be exempted of corrections.

The quasi-equilibrium approximation for the rest of the macroscopic variables is derived in the usual
way. In order to derive the  equation for the velocity, we substitute the local Maxwellian into the
one-particle Liouville equation, and act with the operator $\mu_k=\int v_k \cdot d\vv$ on both the
sides of the equation (\ref{li}). We have:
\begin{eqnarray}
{\partial n u_k\over \partial t}=-{\partial \over
\partial r_k}{nk_BT\over m}-{\partial nu_ku_j\over \partial r_j}.\nonumber
\end{eqnarray}

Similarly,  we derive the equation for the energy density, and the complete system of equations of
the quasi-equilibrium approximation reads (Euler equations):
\begin{eqnarray}
{\partial n\over
\partial t}&=&-{\partial nu_i\over \partial r_i},\label{eu1}\\*
{\partial nu_k\over \partial t}&=& -{\partial \over \partial r_k}{nk_BT\over m}-{\partial nu_ku_j
\over \partial r_j},\nonumber\\ {\partial \varepsilon\over
\partial t}&=&-{\partial \over \partial r_i}\left({5k_BT\over
m}nu_i+u^2nu_i\right).\nonumber
\end{eqnarray}

Now we are going to derive the next order approximation to  the macroscopic dynamics (first order in
the coarse-graining time $\tau$). For the velocity equation we have:
\begin{eqnarray}
R_{nu_k}={1\over 2}\left(\int v_kv_iv_j{\partial^2 f_0\over
\partial r_i\partial r_j}d\vv -\sum_{j}{\partial \phi_{nu_k}\over
\partial M_j}\phi_j\right),\nonumber
\end{eqnarray}
where $\phi_j$ are the corresponding right hand sides of the Euler equations (\ref{eu1}). In order to
take derivatives with respect to macroscopic moments $\{M_0,M_i,M_4\}$, we need to rewrite equations
(\ref{eu1}) in terms of these variables instead of $\{n,u_i,T\}$. After some computation, we obtain:
\begin{eqnarray}
R_{nu_k} ={1\over 2}{\partial \over
\partial r_j}\left({nk_BT\over m}\left[{\partial u_k\over \partial
r_j}+{\partial u_j\over \partial r_k}-{2\over 3} {\partial u_n\over \partial
r_n}\delta_{kj}\right]\right).\label{NS1}
\end{eqnarray}

For the energy we obtain:
\begin{eqnarray}
R_{\varepsilon}={1\over 2}\left(\int v^2v_iv_j{\partial^2 f_0\over
\partial r_i\partial r_j}d\vv -\sum_{j}{\partial \phi_{\varepsilon }\over
\partial M_j}\phi_j\right)={5\over 2}{\partial \over \partial
r_i}\left({nk_B^2T\over m^2 }{\partial T\over \partial r_i}\right).\label{NS2}
\end{eqnarray}

Thus,  we get the system of the Navier-Stokes equation in the following form:
\begin{eqnarray}
{\partial n\over
\partial t}&=&-{\partial nu_i\over \partial r_i},\nonumber\\
{\partial nu_k\over \partial t}&=&-{\partial \over \partial r_k}{nk_BT\over m}-{\partial nu_ku_j
\over \partial r_j}+\nonumber\\* &&{\tau\over 2} {\partial \over \partial r_j}{nk_BT\over
m}\left({\partial u_k\over
\partial r_j}+{\partial u_j\over \partial r_k}-{2\over 3}
{\partial u_n\over \partial r_n}\delta_{kj}\right),\label{NS}\\* {\partial \varepsilon\over
\partial t}&=-&{\partial \over \partial r_i}\left({5k_BT\over
m}nu_i+u^2nu_i\right)+\tau{5\over 2}{\partial \over \partial r_i}\left({nk_B^2T\over m^2}{\partial
T\over
\partial r_i}\right).\nonumber
\end{eqnarray}
We see that kinetic coefficients (viscosity and heat conductivity) are proportional to the
coarse-graining time $\tau$. Note that they are identical with kinetic coefficients as derived from
the Bhatnagar-Gross-Krook model \cite{BGK}  in the first approximation of the Chapman-Enskog method
\cite{Chap} (also, in particular, no bulk viscosity) .

\subsection{Post-Navier-Stokes equations}

Now we are going to obtain the  second-order approximation to the hydrodynamic equations in the
framework of the present approach. We will compare qualitatively the result with the Burnett
approximation. The comparison concerns stability of the hydrodynamic modes near global equilibrium,
which is violated for the Burnett approximation. Though the derivation is straightforward also in the
general, nonlinear case, we shall consider only the linearized equations which is appropriate to our
purpose here.

Linearizing the local Maxwell  distribution function, we obtain:
\begin{eqnarray}
f&=&n_0\left({m\over 2\pi k_BT_0}\right)^{3/2}\left({n\over n_0}+{mv_n\over
k_BT_0}u_n+\left({mv^2\over 2k_BT_0}-{3\over 2 }\right){T\over T_0}\right)e^{-mv^2\over
2k_BT_0}=\nonumber\\* &=&\left\{(M_0+2M_ic_i+\left({2\over 3}M_4-M_0\right)\left(c^2-{3\over
2}\right)\right\}e^{-c^2},\label{mL}
\end{eqnarray}
where we have introduced dimensionless variables: $c_i=v_i/v_T$, $v_T=\sqrt{2k_BT_0/m}$ is the
thermal velocity, $M_0=\delta n/n_0,$ $M_i=\delta u_i/v_T$, $M_4=(3/2)(\delta n/n_0+\delta T/T_0)$.
Note that  $\delta n$, and $\delta T$ determine deviations of these variables from their equilibrium
values, $n_0,$ and $T_0$.

The linearized Navier-Stokes equations read:
\begin{eqnarray}
{\partial M_0\over \partial t}&=&-{\partial M_i\over
\partial r_i},\nonumber\\*
{\partial M_k\over \partial t}&=&-{1\over 3}{\partial M_4\over
\partial r_k}+{\tau\over 4}{\partial \over
\partial r_j }\left({\partial M_k\over
\partial r_j}+{\partial M_j\over \partial r_k}-{2\over 3}
{\partial M_n\over \partial r_n}\delta_{kj}\right),\label{linNS}\\* {\partial M_4\over
\partial t}&=&-{5\over 2}{\partial M_i\over
\partial r_i}+\tau {5\over 2}{\partial^2 M_4\over \partial r_i\partial r_i}.\nonumber
\end{eqnarray}

Let us first compute  the post-Navier-Stokes correction to the velocity equation. In accordance with
the equation (\ref{26}), the first part of this term under linear approximation is:
\begin{eqnarray}
&&{1\over 3!} \mu_k \left({\partial J^*\over \partial f}{\partial J^*\over \partial f} J^*\right) -
{1\over 3!} \sum_{ij} \left( {\partial \phi_k^*\over \partial M_i} {\partial \phi_i^*\over \partial
M_j}\phi_j^*\right)=\nonumber\\* &=&-{1\over 6}\int c_k{\partial^3\over \partial r_i\partial
r_j\partial r_n}c_ic_jc_n\left\{(M_0+2M_ic_i+\left({2\over 3}M_4-M_0\right)\left(c^2-{3\over
2}\right)\right\}e^{-c^2}d^3c\nonumber\\* &+&{5\over 108 }{\partial \over \partial r_i }{\partial^2
M_4\over
\partial r_s\partial r_s}={1\over 6}{\partial \over \partial r_k }\left({3\over 4}
{\partial^2 M_0\over
\partial r_s\partial r_s }-{\partial^2 M_4\over
\partial r_s\partial r_s }\right)
+{5\over 108 }{\partial \over \partial r_k }{\partial^2 M_4\over
\partial r_s\partial r_s}\nonumber\\*
&=&{1\over 8}{\partial \over \partial r_k }{\partial^2 M_0\over
\partial r_s\partial r_s }-{13\over 108}{\partial \over \partial r_k }{\partial^2 M_4\over
\partial r_s\partial r_s }.\label{3u1}
\end{eqnarray}

The part of equation (\ref{26}) proportional to the first-order correction is:
\begin{eqnarray} &-&{1 \over 2 } \sum_j \left(
{\partial \phi_k^*\over \partial M_j} R_j^{(1)}+ {\partial R_k^{(1)}\over
\partial M_j} \phi_j^*\right)={5\over
6}{\partial \over \partial r_k }{\partial^2 M_4\over
\partial r_s\partial r_s }+{1\over 9}{\partial \over \partial r_k }{\partial^2 M_4\over
\partial r_s\partial r_s }.\label{3u2}
\end{eqnarray}
Combining together terms (\ref{3u1}), and (\ref{3u2}), we obtain:
\begin{eqnarray} R_{M_k}^{(2)}={1\over 8}{\partial \over \partial r_k }{\partial^2 M_0\over
\partial r_s\partial r_s }+{89\over 108}{\partial \over \partial r_k }{\partial^2 M_4\over
\partial r_s\partial r_s }.\nonumber
\end{eqnarray}

Similar calculation for the energy equation leads to the following result:
\begin{eqnarray}
&&-\int c^2{\partial^3\over \partial r_i\partial r_j\partial
r_k}c_ic_jc_k\left\{(M_0+2M_ic_i+\left({2\over 3}M_4-M_0\right)\left(c^2-{3\over
2}\right)\right\}e^{-c^2}d^3c +\nonumber\\* &+&{25\over 72}{\partial\over
\partial r_i}{\partial^2 M_i\over
\partial r_s\partial r_s}=-{1\over 6}\left({21\over 4}{\partial\over
\partial r_i}{\partial^2 M_i\over
\partial r_s\partial r_s}+{25\over 12}{\partial\over
\partial r_i}{\partial^2 M_i\over
\partial r_s\partial r_s}\right)=-{19\over 36}{\partial\over
\partial r_i}{\partial^2 M_i\over
\partial r_s\partial r_s}.\nonumber
\end{eqnarray}

The term proportional to the first-order corrections gives:
\begin{eqnarray} {5\over 6}\left({\partial^2\over
\partial r_s\partial r_s}{\partial M_i\over
\partial r_i}\right)+{25\over 4}\left({\partial^2\over
\partial r_s\partial r_s}{\partial M_i\over
\partial r_i}\right).\nonumber
\end{eqnarray}

Thus,  we obtain:
\begin{eqnarray}
R^{(2)}_{M_4}={59\over 9}\left({\partial^2\over
\partial r_s\partial r_s}{\partial
M_i\over \partial r_i}\right).
\end{eqnarray}

Finally, combining  together all the terms, we obtain the following system of linearized hydrodynamic
equations:
\begin{eqnarray}
{\partial M_0\over \partial t}&=&-{\partial M_i\over
\partial r_i},\nonumber\\*
{\partial M_k\over \partial t}&=&-{1\over 3}{\partial M_4\over
\partial r_k}+{\tau\over 4}{\partial \over
\partial r_j }\left({\partial M_k\over
\partial r_j}+{\partial M_j\over \partial r_k}-{2\over 3}
{\partial M_n\over \partial r_n}\delta_{kj}\right)+\nonumber\\*&&\tau^2\left\{{1\over 8}{\partial
\over
\partial r_k }{\partial^2 M_0\over
\partial r_s\partial r_s }+{89\over 108}{\partial \over \partial r_k }{\partial^2 M_4\over
\partial r_s\partial r_s }\right\},\label{3order}\\*
{\partial M_4\over
\partial t}&=&-{5\over 2}{\partial M_i\over
\partial r_i}+\tau {5\over 2}{\partial^2 M_4\over \partial r_i\partial r_i}+{\tau^2}{59\over 9}\left({\partial^2\over
\partial r_s \partial r_s}{\partial
M_i\over \partial r_i}\right).\nonumber
\end{eqnarray}

Now we are in a position to investigate the dispersion relation of this system. Substituting
$M_i=\tilde{M}_i\exp(\omega t+i(\kk,\rr))$ $(i=0,\,k,\,4)$ into equation (\ref{3order}), we reduce
the problem to finding the spectrum of the matrix:
\begin{eqnarray}\left(
\begin{array}{ccccc}
  0 & -ik_x & -ik_y & -ik_z & 0 \\
  -ik_x{k^2\over 8} & -{1\over 4}k^2-{1\over 12}k_x^2 &
-{k_xk_y\over 12}& -{k_xk_z\over 12}& -ik_x\left({1\over 3 }+{89k^2\over 108}\right) \\
  -ik_y{k^2\over 8} & -{k_xk_y\over 12} & -{1\over
4}k^2-{1\over 12}k_y^2&  -{k_yk_z\over 12} & -ik_y\left({1\over 3}+ {89k^2\over 108}\right) \\
  -ik_z{k^2\over 8} &  -{k_xk_z\over 12} &  -{k_yk_z\over 12}
&-{1\over 4}k^2-{1\over 12}k_z^2 & -ik_z\left({1\over 3}+{89k^2\over 108}\right)
\\
  0 &-ik_x\left({5\over 2}+{59k^2\over 9}\right) & -ik_y\left({5\over
2}+{59k^2\over 9}\right) & -ik_z\left({5\over 2}+{59k^2\over 9}\right) & -{5\over 2}k^2
\end{array}\right)\nonumber
\end{eqnarray}

This matrix has five eigenvalues. These real parts of these eigenvalues responsible for the decay
rate of the corresponding modes  are shown in Fig.\ 2 as functions of the wave vector $k$. We see
that {\it all real parts of all the eigenvalues are non-positive for any wave vector}. In other
words, this means that the present system is linearly stable. For the Burnett hydrodynamics as
derived from the Boltzmann or from the single relaxation time Bhatnagar-Gross-Krook model, it is well
known that the decay rate of the acoustic  becomes positive after some value of the wave vector
\cite{Boby,KG2002} which leads to the instability. While the method suggested here is clearly
semi-phenomenological (coarse-graining time $\tau$ remains unspecified), the consistency of the
expansion with the entropy requirements, and especially the latter result of the linearly stable
post-Navier-Stokes correction strongly indicates that it might be more suited  to establishing models
of highly nonequilibrium hydrodynamics.

\subsection{Diffusion in the two-component fluid}

In this example we consider a mixture of the particles of two kinds. We determine microscopic
equations as two independent one-particle Liouville equations. Using our general procedure,
 we obtain a  diffusion behavior of the mixture on
the macroscopic level.

Microscopic equations of motion for the particles of the first and of the second kind are:
\begin{eqnarray}
{\partial f_{\rm 1}\over \partial t}&=&-\vv_1{\partial f_1\over
\partial \rr},\label{VBc}\\*{\partial f_2\over
\partial t}&=&-\vv_2{\partial f_2\over
\partial
\rr}.\label{VBs}
\end{eqnarray}
We denote all variables related to particles of the first kind with index $1$, and with the index $2$
for the second kind, $f_1=f_1(\rr,\vv_1,t)$ and $f_2=f_2(\rr,\vv_2,t)$ are one-particle distribution
functions.

In order to describe hydrodynamics of the system, we introduce the following macroscopic variables:
\begin{eqnarray}
M_1&=&\rho_1=m_1\int f_1 d\vv,\nonumber \\* M_2&=&\rho_2=m_2\int f_2 d\vv,\nonumber\\*
M_i&=&(\rho_1+\rho_2)u_i=m_1\int v_{1i}f_1 d\vv+m_2\int v_{2i}f_1 d\vv,\label{mom}\\* M_T&=&{3\over
2}(n_1+n_2)k_BT=m_1\int {(v_1-u)^2\over 2} f_1d\vv+m_2\int {(v_2-u)^2\over 2} f_2d\vv,\nonumber
\end{eqnarray}
where $\rho_1,\,\rho_2$ are densities; $u_i$ is the i-th spatial component of the average velocity of
the mixture; and $T$ is the temperature.

The quasi-equilibrium distribution functions are:
\begin{eqnarray}
f_1=n_1(\rr)\left(m_1\over 2\pi k_BT(\rr)\right)^{3/2}\exp\left\{-{m_1(v_1-u(\rr))^2\over
2k_BT(\rr)}\right\},\label{fceq}\\*f_2=n_2(\rr)\left(m_2\over 2\pi
k_BT(\rr)\right)^{3/2}\exp\left\{-{m_2(v_2-u(\rr))^2\over 2k_BT(\rr)}\right\}.\label{fseq}
\end{eqnarray}

After the convolution of the system (\ref{VBc})--(\ref{VBs}) with the  operators $\mu_1=\int m_1\cdot
d\vv,$ $\mu_2=\int m_2\cdot d\vv,$ $\mu_{u1}=\int m_1 v_1\cdot d\vv,$ $\mu_{u2}=\int m_2 v_2 \cdot
d\vv,$ $\mu_{T1}=\int m_1 (v_1-u)^2/2\cdot d\vv,$ $\mu_{T2}=\int m_2 (v_2-u)^2/2\cdot d\vv$,  and
after summations, we obtain the system of Euler equations for the binary mixture:
\begin{eqnarray}
{\partial \rho_1\over \partial t}&=&-{\partial \rho_1u_i\over
\partial r_i},\label{Eu1.1}\\*{\partial \rho_2\over \partial
t}&=&-{\partial \rho_2u_i\over
\partial r_i},\label{Eu2.1}\\*
{\partial \rho u_k\over \partial t}&=&-{\partial nk_BT \over
\partial r_k}-{\partial \rho u_ku_i\over \partial
r_i},\label{Eu3.1}\\* {3\over 2}{\partial nk_BT\over
\partial t}&=&-{3\over 2}{\partial u_ink_BT\over
\partial r_i}-nk_BT{\partial u_i\over \partial r_i}.\label{Eu4.1}
\end{eqnarray}

Let us now calculate the dissipative correction for the density equations. For the equation
(\ref{Eu1.1}) we obtain:
\begin{eqnarray}
R_1&=&{1\over 2}\left({\partial^2n_1k_BT\over
\partial r^2}+{\partial^2 \rho_1 u_iu_j\over \partial r_i\partial
r_j}\right)\nonumber\\* &-&{1\over 2}{\partial \over
\partial r_k }\left({M_k\over \rho}{\partial \over
\partial r_j }\left({M_j\rho_1\over \rho}\right)-
{M_k\rho_1\over \rho^2}{\partial \over
\partial r_j }\left({M_j\rho_1\over
\rho}\right)\right)\nonumber\\*&-&{1\over 2}{\partial \over
\partial r_k }\left({M_k\rho_2\over \rho^2}{\partial \over
\partial r_j }\left({M_j\rho_2\over \rho}\right)\right)-
{1\over 2}{\partial \over \partial r_k}\left\{{\rho_1\over \rho}{\partial nk_BT\over \partial r_k
}+{\rho_1\over \rho}{\partial nu_ju_k\over
\partial r_j }\right\}.\label{Rc}
\end{eqnarray}

Substituting (\ref{mom}) into (\ref{Rc}), and performing  similar to the above calculations for the
equation (\ref{Eu1.1}), we arrive at the diffusion equations for the binary mixture under a
simplifying assumption $T=\textrm{const}$, $\uu=0$:
\begin{eqnarray}
{\partial \rho_1\over \partial t}&=&{\tau k_BT\over 2}\left\{{\partial^2n_1\over \partial r_k\partial
r_k}-{\partial \over
\partial r_k }\left({\rho_1\over \rho}{\partial n\over
\partial r_k } \right)\right\},\label{Ei1}\\*{\partial \rho_2\over
\partial t}&=&{\tau k_BT\over 2}\left\{{\partial^2 n_2\over \partial
r_k\partial r_k}-{\partial \over
\partial r_k }\left({\rho_2\over \rho}{\partial n\over
\partial r_k }\right)\right\}.\label{Ei2}
\end{eqnarray}
 Diffusion coefficient
$\tau k_BT/2m$ coincides with Einstein diffusion coefficient, where $\tau$ has meaning of the average
relaxation time.

\section{Hydrodynamic equation for the fluid with long-range
interaction}

In this section we derive equations of hydrodynamics from the nonlinear Vlasov equation. This example
is also interesting from the methodological point of view. Usual methods of reduction of the
description are applicable mostly  to a  linear microscopic dynamics \cite{Rob,Zub,Grab}. These
methods are based on a formal solution to the microscopic equation of motion presented in the
exponential form. This is not directly possible for nonlinear microscopic models. Since we avoid
integrating microscopic equations of motion, our approach is immediately applicable to nonlinear
systems without any modifications.

Let the microscopic dynamics be given by the Vlasov equation (\ref{V1}). Choosing again the usual
hydrodynamic fields for the
 macroscopic variables (\ref{M}),  we
obtain the system of Euler equations enriched by the mean-field terms:
\begin{eqnarray}
{\partial n\over
\partial t}&=&-{\partial nu_i\over \partial r_i},\nonumber
\\*
{\partial nu_k\over \partial t}&=&-{\partial \over \partial r_k}{nk_BT\over m}-{\partial nu_ku_j
\over \partial r_j}+nF_k,\label{mean0}\\* {\partial \varepsilon\over
\partial t}&=&-{\partial \over \partial r_i}\left({5k_BT\over
m}nu_i+u^2nu_i\right)+2nu_iF_i,\nonumber
\end{eqnarray}
where $F_i$ is the $i-$th spatial component of the mean-field force given by equation  (\ref{FF}).

Following the same route  as in section (\ref{ChH}),  we compute term by term the  correction for
(\ref{mean0}). The first term in the brackets in equation (\ref{Rk1}) is proportional to:
\begin{eqnarray}
{\partial J^*\over \partial f}J^*&=&v_iv_j{\partial^2f_0\over
\partial r_i\partial r_j}+v_i{\partial \over \partial
r_i}\left({\partial f_0\over \partial v_j}F_j\right)+{\partial f_0\over \partial
v_i}\Psi_i\nonumber\\* &+&F_i{\partial \over
\partial v_i}\left(v_j{\partial f_0\over \partial
r_j}\right)+F_i{\partial \over \partial v_i}\left({\partial f_0\over \partial
v_j}F_j\right),\label{HV}
\end{eqnarray}
where $\Psi$ is given by equation(\ref{Pmf}).

In order to calculate the mean-field terms in the velocity equation, we act by the operator $\int v_k
\cdot d\vv$ on equation  (\ref{HV}). We obtain:
\begin{eqnarray}
&&-2\left( {\partial nu_i\over \partial r_i}F_k+{\partial nu_k\over
\partial r_i}F_i\right)-(nu_i\delta_{jk}+nu_k\delta_{ij}){\partial
F_j\over \partial r_i}\nonumber\\* &&-n\Psi_k+{\partial nu_k\over
\partial r_i}F_i.\label{int2}
\end{eqnarray}

The second part of equation (\ref{Rk1}) is:
\begin{eqnarray}
\sum_j{\partial \phi_{u_i}\over \partial M_j}\phi_j=-F_jM_0{\partial M_i\over \partial
r_j}-M_j{\partial F_i\over \partial r_j}-F_i{\partial M_j\over \partial r_j}-{M_i\over M_0}{\partial
F_jM_0\over
\partial r_j}-F_i{\partial M_j\over \partial r_j},\nonumber
\end{eqnarray}
where $M_i$ correspond to (\ref{M}). Rewriting this expression in terms of the variables $n,\uu$ and
$T$, and combining the result with equation (\ref{int2}), we obtain:
\begin{eqnarray}
R^{(1)}_{nu_k}=n\Psi_k.\nonumber
\end{eqnarray}

Now let us calculate the correction to  the energy equation (\ref{NS}) in the presence of the
mean-field interaction. Action of the operator $\int v^2 \cdot d\vv$ on equation (\ref{HV}) gives:
\begin{eqnarray}
&&\left(-5{\partial \over \partial r_j}\left({k_BTn\over m}\right)-2u_iu_j{\partial n\over \partial
r_i}-u^2{\partial n\over
\partial r_j}\right)F_j-\nonumber\\* &&\left(2nu_i{\partial u_j\over
\partial r_i}
-2nu_i{\partial u_i\over \partial r_j}-2nu_j{\partial u_i\over
\partial r_i}\right)F_j+\nonumber\\*
&&\left(-5{nk_BT\over m}\delta_{ij}-2nu_iu_j-u^2n\delta_{ij}\right){\partial F_j\over
\partial r_i}\nonumber\\*
&&-2nu_i\Psi_i+ \left(3{\partial \over \partial r_i}\left({k_BTn\over m }\right)+{\partial nu^2\over
\partial r_i}\right)F_i+2nF_iF_i.\nonumber
\end{eqnarray}

The differential term for the energy density equation gives:
\begin{eqnarray}
\sum_i{\partial \phi_{\varepsilon}\over \partial M_i}\phi_i&=&-{M_4\over M_0}F_i{\partial M_0\over
\partial r_i}-M_4{\partial F_i\over
\partial r_i}-{2\over 3}{\partial \over
\partial r_i}\left(F_i\left(M_4-{M_nM_n\over
M_0}\right)\right)\nonumber\\&-&2F_iM_j{\partial\over \partial r_j}\left({M_i\over M_0}\right)-
2F_jM_i{\partial\over
\partial r_j}\left({M_i\over M_0}\right)-2{M_iM_j\over
M_0}{\partial F_i\over
\partial r_j}\nonumber\\ &-&F_i{\partial \over \partial
r_i}\left(M_4-{M_nM_n\over M_0}\right)+{F_i\over M_0}\left(M_4-{M_nM_n\over M_0}\right){\partial
M_0\over \partial r_i}\nonumber\\* &-&{2M_jF_j\over M_0}{\partial M_i\over
\partial r_i}-F_i{2\over 3}{\partial\over
\partial r_i}\left(M_4-{M_nM_n\over
M_0}\right)\nonumber\\* &-&2F_iM_j{\partial\over
\partial r_j}\left(M_i\over M_0\right)-{2F_iM_i\over M_0}{\partial M_j\over
\partial r_j}.\nonumber
\end{eqnarray}

Thus, we obtain the first-order correction to the energy equation,
\begin{eqnarray}
R^{(1)}_{\varepsilon}={6k_BT\over m}F_i{\partial n\over
\partial r_i}+2F_iu_iu_j{\partial n\over \partial r_j}-2nu_i\Psi
_i+2nF_iF_i,\nonumber
\end{eqnarray}

Finally, we arrive at  the following system of hydrodynamic equations with the  the mean-field
interaction:
\begin{eqnarray}
{\partial n\over
\partial t}&=&-{\partial nu_i\over \partial r_i},\nonumber
\\*
{\partial nu_k\over \partial t}&=&-{\partial \over \partial r_k}{nk_BT\over m}-{\partial nu_ku_j
\over \partial r_j}+nF_k+\nonumber\\* &&{\tau\over 2} {\partial \over
\partial r_j}{nk_BT\over
m}\left({\partial u_k\over
\partial r_j}+{\partial u_j\over \partial r_k}-{2\over 3}
{\partial u_n\over \partial r_n}\delta_{kj}\right)-{\tau n\over 2} \Psi_k,\label{Hmf}\\* {\partial
\varepsilon\over
\partial t}&=-&{\partial \over \partial r_i}\left({5k_BT\over
m}nu_i+u^2nu_i\right)+2nu_iF_i+\nonumber\\&&\tau{5\over 2}{\partial \over \partial
r_i}\left({nk_B^2T\over m^2}{\partial T\over
\partial r_i}\right)+\tau{3k_BT\over m}F_i{\partial n\over
\partial r_i}+\tau F_iu_iu_j{\partial n\over \partial r_j}-\nonumber\\
&&\tau nu_i\Psi_i+\tau nF_iF_i,\nonumber
\end{eqnarray}

Equations (\ref{Hmf}) are the general form of the hydrodynamic equations of a simple fluid with the
mean-field interaction. Examples of the system for which this result may be relevant can be found in
studies of electron transport in the various media \cite{Bal}, as well as in description of
non-Newtonian fluids \cite{coll}. For each particular case the interaction potential
$\Phi(|\rr-\rr'|)$ has its specific form, and leads to the corresponding hydrodynamics of the system.

\section{Conclusion}

In this  paper the formalization of Ehrenfest's approach to irreversible dynamics is given in
details. This method allows one to derive macroscopic equations of motion on the basis of the
microscopic dynamics and the very transparent coarse-graining procedure. The method is applicable to
both reversible as well as to the irreversible microscopic dynamics, independently of whether it is
linear or not. We have presented a set of examples demonstrating how this method is applied to
various situations.

The most interesting continuation of this approach is, of course, how to specify in a sensible and
practical way the coarse-graining time $\tau$ in order to make the modelling parameter-free. This
requires is a subject of our current studies (see \cite{GK2002a,GK2002b,GK2002c}).

Finally, whereas we have focused on the application of our formalism to the entropy-conserving
microscopic dynamics, it should be mentioned that it is applicable also to constructing slow
invariant manifolds of dissipative systems. In particular, when applied to the Boltzmann equation,
the result is equivalent at $\tau\to\infty$ to the exact Chapman-Enskog solution \cite{GK2002a}.

\newpage

\newpage

\begin{figure}
\vspace*{5cm}
\includegraphics{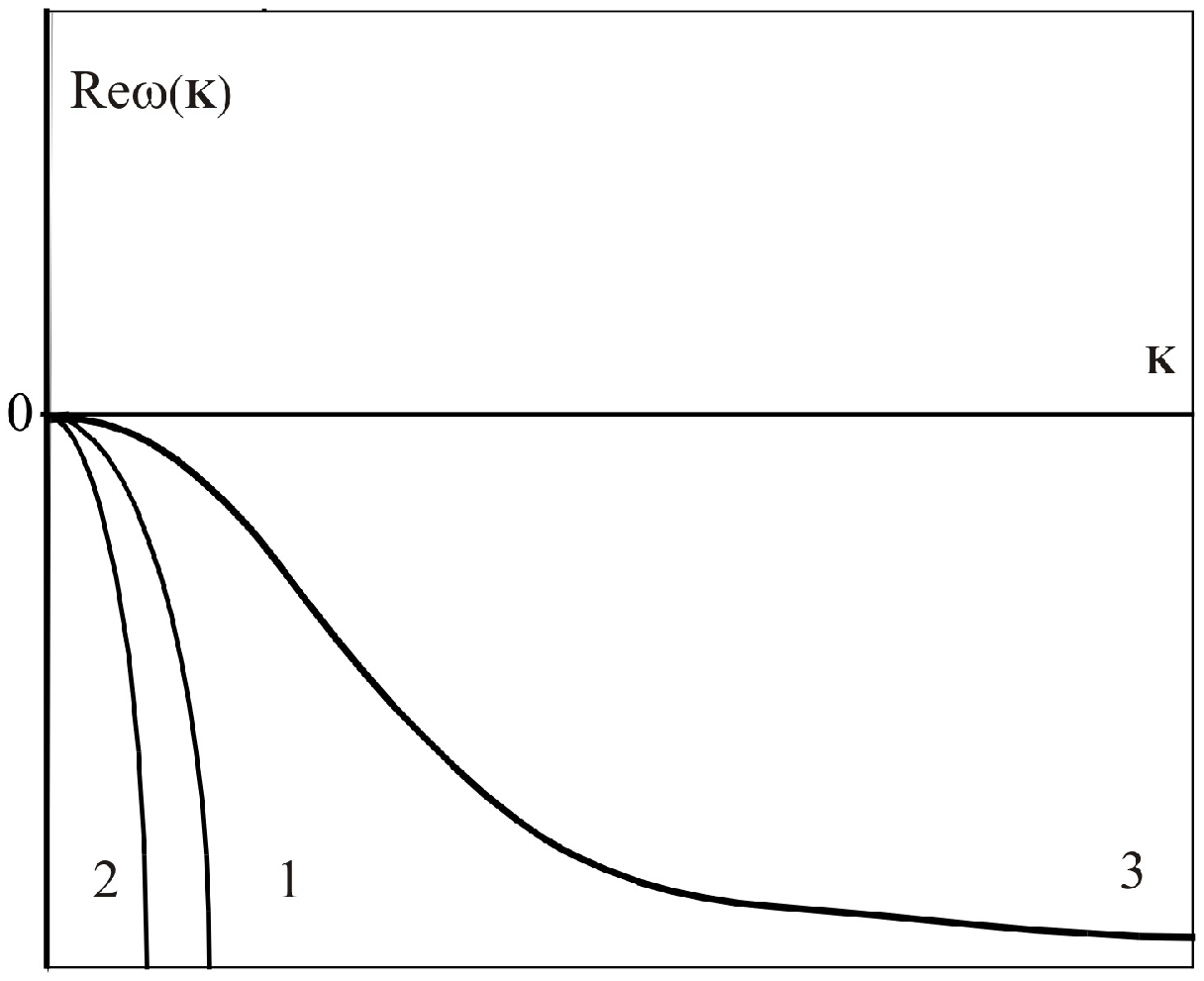}
\caption{Attenuation rates  of various modes of the post-Navier-Stokes equations as functions of the
wave vector. Attenuation rate of the twice degenerated shear mode is curve 1. Attenuation rate of the
two sound modes is curve 2. Attenuation rate of the diffusion mode is curve 3.} \label{Fig2}
\end{figure}


\begin{thebibliography}{99}

\bibitem{Bol} L.\ Boltzmann, {\it Lectures on the Theory of Gases}, University of
California Press, (1964)

\bibitem{Ehr}P.\ Ehrenfest and T.\  Ehrenfest,
{\it Encyklop\"{a}die der Mathematischen Wissenschaften,} Bd.\ IV 2, II, H.\ 6 (1911) [Reprinted in:
P.\ Ehrenfest, {\it Collected Scientific Papers,} North-Holland, Amsterdam, (1959)

\bibitem{See}{\it Studies in Statistical Mechanics}, edited by E.\
W.\ Montroll and J.\ L.\ Lebowitz, North-Holland, (1981), Vol. 9.

\bibitem{Rio}J.\ L.\ del R{\scriptsize{\'{1}}}o-Correa and L.\ S.\
Gars{\scriptsize{\'{1}}}a-Col{\scriptsize{\'{1}}}n,  Phys. Rev. E {\bf 48}, 819 (1993).

\bibitem{GKOT} A.\ N.\ Gorban, I.\ V.\ Karlin, H.\ C.\ \"{O}ttinger, and
L.\ L.\ Tatarinova,  Phys. Rev. E  {\bf 63}, 066124 (2001).

\bibitem{Boby} A.\ V.\ Bobylev, Dokl. Akad. Nauk (SSSR) {\bf 262}, 71
(1982), [Sov.\ Phys.\ Dokl. {\bf 27}, 29 (1982)].

\bibitem{Zub} D.\ Zubarev, V.\ Morosov, and G.\ R\"{o}pke, {\it Statistical
Mechanics of Nonequilibrium Processes,} Akademie Verlag, Berlin, (1996), Vol. 1.

\bibitem{Rob}B.\ Robertson, Phys. Rev. {\bf 44}, 151 (1966).

\bibitem{Grab}H.\ Grabert, {\it Projection Operator Techniques in
Nonequilibrium Statistical mechanics,} Springer, Berlin, (1982).

\bibitem{Jan}E.\ T.\ Jaynes, Phys. Rev. {\bf 106}, 620 (1957); {\bf
108}, 171 (1957).

\bibitem{Gor} A.\ N.\ Gorban, {\it Equilibrium Encircling,} Nauka,
Novosibirsk, (1984).


\bibitem{Lewis}R.\ M.\ Lewis, {\it J.\ Math.\ Phys.} {\bf 8}, 1448
(1967).

\bibitem{RG}O.\ Pashko and Y.\ Oono, {\it Int.\ J.\ Mod.\ Phys. B} {\bf 14},
555 (2000).

\bibitem{Lan} E.\ M.\ Lifschitz and L.\ P.\ Pitaevsky {\it Physical Kinetic,}
Pergamon Press, Oxford, (1980).

\bibitem{Res}P.R\'esibois and  M. De Leener,
{\it Classical Kinetic Theory of Fluids,} Wiley, NY, (1977).

\bibitem{Chap} S.\ Chapman and T. \ G. \ Cowling. {\it The Mathematical Theory
of Non-uniform Gases,} Cambridge University Press, Cambridge, (1970).




\bibitem{GKIO} A.\ N.\ Gorban, I.\ V.\ Karlin, P.\ Ilg, and H.\ C.\
\"{O}ttinger, J. Non-Newtonian Fluid Mech. {\bf 96}, 203 (2001).



\bibitem{HCO} H.\ C.\ \"{O}ttinger,  {\it Stochastic Processes in Polymeric Fluids.
Tools and Examples for Developing Simulation Algorithms} (Springer, Berlin etc, 1996).


\bibitem{Bal} R.\ Balescu. {\it Transport Processes in Plasmas,
Classical Transport Theory,} North-Holland, Amsterdam, (1988), Vol. 1.


\bibitem{KG2002}I.\ V.\ Karlin and A.\ N.\ Gorban, Ann.\ Phys. (Leipzig)
 {\bf 11}, 783 (2002).

\bibitem{BGK}P. L. Bhatnagar, E. P. Gross, and M. Krook,
Phys. Rev. {\bf 94}, 511 (1954).

\bibitem{coll} A.\ N.\ Gorban, I.\ V.\ Karlin, and
L.\ L.\ Tatarinova (unpublished).

\bibitem{GK2002a}A.\ N.\ Gorban and I.\ V.\ Karlin,
Phys.\ Rev.\ E {\bf 65}, 026116 (2002).

\bibitem{GK2002b}A.\ N.\ Gorban and I.\ V.\ Karlin,
Rev.\ Mex.\ F\'{\i}s. {\bf 48 S1}, 238 (2002).

\bibitem{GK2002c}A.\ N.\ Gorban and I.\ V.\ Karlin,
in: {\it  Developments in Mathematical and Experimental Physics, Volume C: Hydrodynamics and
Dynamical Systems}, Eds. A. Macias, F. Uribe and E. Diaz (Kluwer, New York, 2003).

\end{thebibliography}
\end{document}